\documentclass{mypaper}
\usepackage{graphicx}
\usepackage{rotate}
\usepackage{rotating}
\usepackage{relsize}
\usepackage{lineno}
\usepackage{slashed}
\usepackage{url}
\usepackage{amsmath}
\usepackage{longtable}
\begin{document}

%
\begin{titlepage}
\title{Snowmass Energy Frontier Simulations using the Open Science Grid 
\\ 
A Snowmass 2013 whitepaper.
}
 \begin{Authlist}
Aram Avetisyan
 \Instfoot{boston}{Boston University, Boston, USA}
Saptaparna Bhattacharya, Meenakshi Narain
 \Instfoot{brown}{Brown University, Providence, USA}
Sanjay Padhi
 \Instfoot{ucsd}{University of California, San Diego, USA}
Jim Hirschauer, Tanya Levshina, Patricia McBride, Chander Sehgal, Marko Slyz
 \Instfoot{fnal}{Fermi National Accelerator Lab, Batavia, USA}
Mats Rynge
 \Instfoot{isi}{Information Sciences Institute, Marina del Rey, USA}
Sudhir Malik 
 \Instfoot{UNL}{University of Nebraska, Lincoln, USA}
John Stupak III
\Instfoot{purdue}{Purdue University Calumet, Hammond, USA}
\end{Authlist}

\begin{abstract}
Snowmass is a US long-term planning study for the high-energy community by
the American Physical Society's Division of Particles and Fields. For its simulation studies, 
opportunistic resources are harnessed using the Open Science Grid infrastructure.
Late binding grid technology, GlideinWMS, was  used for distributed scheduling of the simulation 
jobs across many sites mainly in the US. The pilot infrastructure also uses the Parrot mechanism to 
dynamically access CvmFS in order to ascertain a homogeneous environment across the nodes. 
This report presents the resource usage and the storage model used for 
simulating large statistics
Standard Model backgrounds needed for Snowmass Energy Frontier studies.
\end{abstract}
\end{titlepage}

\section{Introduction}
The current experimental particle physics projects consist of thousands of physicists from across the globe.
These large scale international collaborations have big budgets. Their future endeavors and road-map
to potential discoveries requires careful long term planning. 
The Snowmass effort epitomizes the work by the high-energy physics community in the United States (US)
to collaborate and plan out a common future keeping in mind the long-term physics aspirations 
and leadership of the US in the field of particle physics. 
Via this effort, the community plans to  communicate the opportunities for 
discoveries in high-energy physics to the broader scientific 
community and the US government. 

For simulation of large backgrounds as expected from $pp$ collisions at
$\sqrt{s} = 14, 33$ and $100$~TeV, opportunistic resources are harnessed using the Open Science Grid 
infrastructure (OSG)~\cite{osg}. OSG facilitates distributed computing for scientific research in the US.
It is comprised of about 126 institutions with 121 recently active sites.

In order to evaluate the physics discovery potential from the recently observed Higgs boson as well as new physics searches, 
large-statistic Standard Model (SM) background Monte Carlo (MC) samples are needed to model the data expected from an integrated luminosity corresponding to  300 fb$^{-1}$ expected  after the long shutdown-1 (LS1) and 3000 fb$^{-1}$ for High Luminosity (HL) Large Hadron Collider (LHC)~\cite{LHC}. 
This would require simulating billions of events calculated using exact matrix element calculations with parton shower. 

The LHC Physics Center (LPC) at Fermilab is a hub of US physicists 
collaborating on the CMS experiment at the LHC and 
plays a key role in various physics analyses and discovery process within the experiment. 
The LPC  has a prime role to play in the Snowmass process. It has vast computing resources and 
in-house experts on physics, experimentation, simulation, software and computing tools.
Being located at Fermilab, the LPC members have access to the OSG personnel and infrastructure. 
The snowmass efforts at LPC represents a collaborative effort carried out by a group of members 
drawn from CMS and ATLAS experiments as well as theorists. This group has taken the lead 
in a very short time to successfully put together an infrastructure to generate the billions of
events needed by the studies for snowmass and help drive the future course of the US HEP program. 
The dedicated team at the LPC has been able to generate a huge amount of Monte Carlo data based 
on simulating the response  of a generic snowmass detector~\cite{snowmassPeformance} to the 
passage of particles produced from proton-proton collisions at several TeV center-of-mass energies. 
With billions of simulated data events for standard model physics processes and new physics
signals of interests for the future, the LPC team and the OSG team have together enabled
studies which help define the future vision of the US program for the Energy Frontier.

In section~\ref{sec:sim}, we discuss the overview and requirements of 
the simulations followed by a detailed  description of the mechanism used to harness 
resources distributed across various sites in section~\ref{sec:infrastructure}. 
In sections~\ref{sec:utilization} and~\ref{sec:storage}, we outline the 
OSG resources used and storage model used for this exercise, 
followed by a summary in section~\ref{sec:summary}.

\section{Simulation Overview and Production Tools}
\label{sec:sim}

The Snowmass LPC group has been running two types of jobs. 
The first set of jobs, ``{\it event generation}'', 
runs matrix-element package, {\sc MadGraph}~\cite{madgraph} to generate 
physics processes from the proton-proton collisions (or ``events'') expected from the LHC. The output from these jobs then serves as 
input to the second class of jobs. In this second set of jobs, ``{\it detector simulation and reconstruction}'', 
the generated events are processed  serially by packages named 
{\sc Bridge}~\cite{bridge}, {\sc Pythia}~\cite{pythia}, 
and {\sc Delphes}~\cite{delphes}, with the resulting output transferred to the storage locations. 
These outputs contain data similar to the data we get from actual proton-proton collisions: collections 
of leptons, jets and other particles, along with their characteristics (energy, momentum, etc). 

The first set of ``{\it event generation}'' jobs needs instructions regarding the type of events to simulate. 
These instructions come in the form of a ``gridpack,'' and given that their size is O(10MB)  they are transferred by HTCondor with the jobs. 
The output of the {\it event generation} stage is in Les Houches Accord (LHE) 4-vector event format, with a size of O(10MB), 
and is  transferred back to OSG-XSEDE via HTCondor.

The second set of ``{\it detector simulation and event reconstruction}'' jobs  simulate 
the accelerator conditions, detector performance, and reconstruction of the particles (or objects) in an 
event. Simulation of the   accelerator conditions with around 50 to 140 extra interactions per bunch crossing,  termed as pile-up interactions, is one of the challenges of this step. To simulate these conditions, we need access to a  large ``minimum bias'' file, which is O(1GB). In an earlier iteration of our recipe 
the file size was significantly large, almost 50GB. This is too large to transfer with each job and 
hence we developed a solution which leads to much less data transfer. This
solution is detailed in the next section.

\section{OSG Distributed Computing Infrastructure and Resources}
\label{sec:infrastructure}

Job submission was done through HTCondor using
GlideinWMS \cite{glideinwmsA}. GlideinWMS essentially starts a job,
called a {\it pilot job}, on each worker node. The pilot job advertises
its existence to the wider system and eventually runs
one or more science jobs depending on their length. GlideinWMS
is the standard submission system that is recommended
to new users.

We shipped the science applications with the user job
but did not necessarily ship the software dependencies of these
applications. For the latter we used the following methods:
\begin{enumerate}
\item Use the software in the CMSSW stack \cite{cmssw} that is
   generally available at sites through CvmFS \cite{cvmfs}.
   CvmFS also enables caching and thus is efficient and scalable.

  \item Use Parrot \cite{parrot} to access the CMSSW software instead
   of directly accessing CvmFS. A Parrot client can be shipped with the
   job so this allows jobs to run at sites that doesn't necessarily have CvmFS installed.
  
\item Ship the needed software dependencies with each job.
   This first involves determining which of the application's
   software dependencies aren't installed on all worker nodes.
   Those dependencies are put into a tar
   archive that HTCondor moves to the worker node. Finally 
   the jobs have to set up the right environment variables like LD\_LIBRARY\_PATH.
  \end{enumerate}
   
The minimum bias file is pre-staged to Storage Elements at 10 different grid sites using the OSG Public Storage
Service built on top of iRODS~\cite{irods, osgpublicstorage}. When a job starts, it checks for the presence of the minimum bias file
in a temporary directory associated with the Glidein, and uses it if available. In absence of the file, the job uses
the OSG Public Storage to get the file from one of the grid sites it is staged at, and replicates it into the Glidein
temporary (temp) directory. Thus, the data transfer is spread across several different sites, and the file is reused
when possible. Some sites do not allow installing an irods-osg client on pilot start up, so on those sites we use
“SRM”~\cite{srm} to directly replicate the inputs with a hardcoded “SURL” from either FNAL, UNL, or BNL.
The OSG Public Storage handles users and resources quotas and is capable of uploading and downloading files to
the OSG resources on user request. It provides a unified ”namespace” for all the stored files. The user can request
data without knowing the exact location of the file. This significantly simplifies the task of accessing the data.
SRM~\cite{srm} is a protocol for accessing large disk or tape arrays. SRM was particularly useful for Snowmass because
it can queue up transfer requests if several of them arrive at once, this reduces the chances of overwhelming the underlying server.

\section{Utilization of OSG resources}
\label{sec:utilization} 
   
\begin{figure}[!ht]
\begin{center}
\includegraphics[width=0.9\textwidth]{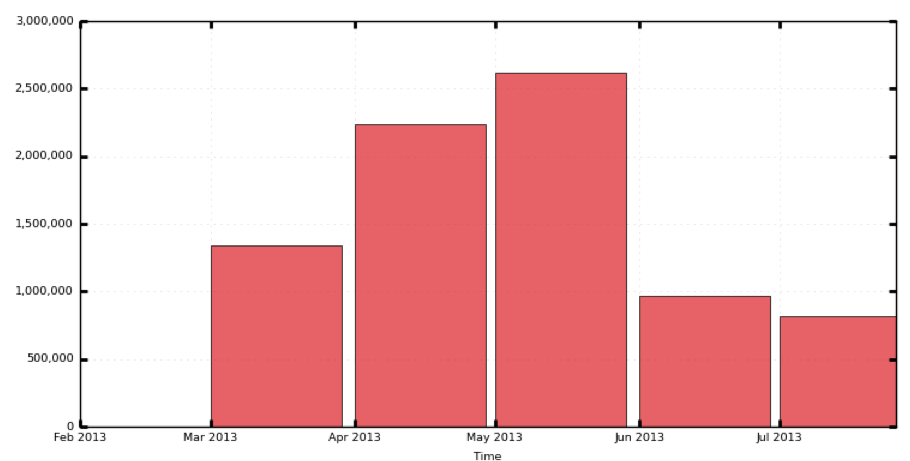}
\caption{\label{fig:hours} Computing hours (walltime) used by the Snowmass project on the OSG.}
\end{center}
\end{figure}

\begin{figure}[!ht]
\begin{center}
\includegraphics[width=0.9\textwidth]{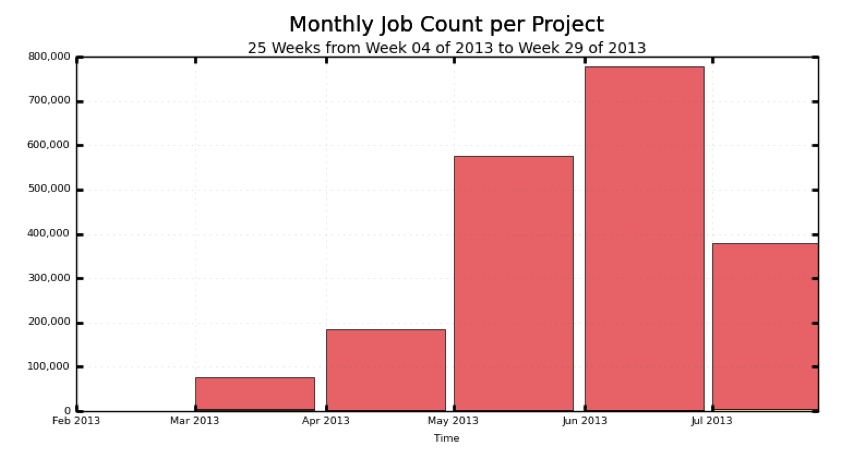}
\caption{\label{fig:jobs} Number of jobs as a function of time used by the Snowmass project on the OSG.}
\end{center}
\end{figure}

 Resources are harnessed using an OSG submit host at Indiana University. Jobs were submitted to all sites, mostly in
the US. The top contributors are shown in Table~\ref{tab:sites}. The resource utilization information was obtained
from the OSG Gratia accounting service~\cite{gratia}. Since February a total of 7.8 million hours has been used by the
Snowmass group. Figure~\ref{fig:hours} demonstrates the usage of computing hours, specifically walltime, on a monthly basis.
More than 100K CPU hours were used on a significant number of days starting from March to May. The phase of operations
prior to May, phase 1, yielded about 1.5 Billion LHE events, and 0.5 Billion output events from {\sc Delphes}.
The LHE event generation scheme used prior to mid May was processing intensive. The “inclusive event
generation” scheme was later changed to the “ST binned weighted event generation”~\cite{snowmassBkg} scheme and led to much
reduced processing times.
Figure~\ref{fig:jobs} shows a monthly summary of the complted number of jobs. About 14K jobs were complted per day, with a peak of around
70-80K jobs run during the last week of May 2013 and first week of June 2013. These were due to the LHE 
event generation at $\sqrt{s}$=14 TeV in May and subsequent {\sc Delphes} processing in early June. The {\sc Delphes} jobs consumed
less processing time, and hence do not contribute as much of a load in the distribution of monthly hours used (see also
Figure~\ref{fig:hours}).

\begin{table}[!ht]
\begin{center}
\caption{The number of hours contributed by the top 15 OSG computing sites.}
\begin{tabular}{|r|r|}
\hline
Site & Wall duration (hours) \\
\hline
USCMS-FNAL-WC1       & 3411395 \\
Nebraska             &  775651 \\
UCSDT2               &  677377 \\
Firefly              &  367587 \\
Tusker               &  338262 \\
CIT\_CMS\_T2           &  327662 \\
MWT2                 &  303514 \\
GridUNESP\_CENTRAL    &  205213 \\
FNAL\_FERMIGRID       &  185868 \\
SPRACE               &  169428 \\
UFlorida-SSERCA      &  159302 \\
UCD                  &  118339 \\
UConn-OSG            &  117724 \\
TTU-ANTAEUS          &  113602 \\
SMU\_HPC              &   72257 \\
Other (17 sites)      & 465426 \\
\hline
\end{tabular}
\label{tab:sites} 
\end{center}
\end{table}

\section{Storage Model}
\label{sec:storage}

The outputs of the second stage contain collections of objects similar to what we 
expect from real proton-proton collisions. The size of these collections grow with the additional pile-up 
interactions and is $\approx$5-7 kB with  zero PU, and $\approx$15-20 kB with 140 PU.
The second stage outputs from the {\sc Delphes} package are in ROOT~\cite{rootA} format. 

The output data can get stored on disks in one of three places:
the USCMS Tier 1 dCache at Fermilab, the Hadoop file system
at University of Nebraska Lincoln, or the dCache server at
the ATLAS Tier 1 at Brookhaven National Lab. All of these are
set up to require a certificate to read or write. To help with
storage authorization we set up a Snowmass group as part of
the OSG VO and added all the users who needed write access
to this group.

In addition, the UNL storage is accessible
for read-only access through a web interface and XRootD~\cite{XRootD}. Therefore the samples at UNL 
are used by the entire HEP community without being restricted to a given LHC virtual organization (VO). 

The samples' storage locations are listed in Table~\ref{tab:storage}. Table~\ref{tab:dataload} shows the volume of data transferred to and from the Fermilab dCache and the UNL storage. 
The LHE event files are archived at LPC, the simulation output datasets at FNAL and UNL storage elements for 
long term storage.

\begin{table}[!ht]
\begin{center}
\caption{Storage Location of Generated Event Samples (in {\sc Delphes} output ROOT format).}
\begin{tabular}{l|l} \hline\hline
\multicolumn{2}{l}{\bf Events at $\sqrt{s}$=14 and, 33 TeV} \\ \hline
{\bf UNL Site:} & \\ 
PU $<\mu>$=0 &  \url{http://red-gridftp11.unl.edu/Snowmass/HTBinned/Delphes-3.0.9.1/NoPileUp} \\
PU $<\mu>$=50 & \url{http://red-gridftp11.unl.edu/Snowmass/HTBinned/Delphes-3.0.9.1/50PileUp} \\
PU $<\mu>$=140 &\url{http://red-gridftp11.unl.edu/Snowmass/HTBinned/Delphes-3.0.9.1/140PileUp} \\  
PU $<\mu>$=0 &  \url{http://red-gridftp11.unl.edu/Snowmass/HTBinned/Delphes-3.0.9.2/NoPileUp} \\
PU $<\mu>$=50 &  \url{http://red-gridftp11.unl.edu/Snowmass/HTBinned/Delphes-3.0.9.2/50PileUp}\\
PU $<\mu>$=140 &  \url{http://red-gridftp11.unl.edu/Snowmass/HTBinned/Delphes-3.0.9.2/140PileUp}\\ \hline
{\bf BNL Site: }& \url{https://dcdoor10.usatlas.bnl.gov:2881/pnfs/usatlas.bnl.gov/osg/snowmass}\\ \hline
{\bf FNAL CMSLPC :} & \url{/pnfs/cms-pg/11/store/user/snowmass} \\ \hline
\hline
\end{tabular}
\label{tab:storage}
\end{center}
\end{table}
 
\begin{table}[!ht]
\begin{center}
\caption{Amount of data transfer  to/from the Fermilab CMS dCache and the UNL storage.}
\begin{tabular}{|l|c | c |} \hline
Month &	Fermi dCache (TB) & 	UNL (TB)\\
June &	65.0 &	46.4\\
May &	12.4 &	5.2\\
April& 	189.7 &	10.8\\
March &	1.1 &	0.0\\
Total &	268.3& 	62.5\\ \hline
\end{tabular}
\label{tab:dataload}
\end{center}
\end{table}

\section{Summary}
\label{sec:summary}

New methods and tools were necessary to simulate large physics backgrounds for future endeavors and road-maps by 
the US high-energy physics community. Opportunistic resources were harnessed at a large scale using the Open Science 
Grid infrastructure. The results of the studies, involving Higgs, Top, and New Physics searches, etc., based on simulated events
enabled by the OSG resources, will be presented in the Snowmass workshop at Minneapolis, July 29 - Aug 6, 2013.
These studies will provide guidance on the future direction of the Energy Frontier to the scientific community. 
The help from OSG and FNAL LPC was crucial in the success of this program.

\section{Acknowledgments}
This research was done using resources provided by the Open Science Grid, which is
supported by the National Science Foundation and the U.S. Department of Energy's Office of Science.
The storage elements from USCMS T1 at FNAL, the Holland Computing Center at UNL, and the Atlas T1 at BNL were
used for the studies. We are thankful for help from the members of the AAA~\cite{AAAref} project and OSG teams with 
the Parrot integration, the CILogon and with distributed GlideinWMS usage and support.

\end{document}